\documentclass[12 pt,twoside,aps,prd,amsmath,amssymb,
tightenlines,showkeys,showpacs,eqsecnum]{revtex4}
\usepackage{epsfig}
\usepackage{bm}

\usepackage{hyperref}

\newcommand {\ve}{\varepsilon}

\def \myfigures #1#2#3#4#5#6#7#8
{\begin{figure}[ht]
    \begin{center}
        \includegraphics[width=#2 \textwidth]{#1.eps}
        \hfill
        \includegraphics[width=#6 \textwidth]{#5.eps}
        \parbox[t]{#4\textwidth}{\caption {#3}\label{#1}}
        \hfill
        \parbox[t]{#8\textwidth}{\caption {#7}\label{#5}}
    \end{center}
\end{figure} }

\def \myfigs #1#2#3#4#5#6#7#8
{\begin{figure}[ht]
    \begin{center}
        \includegraphics[width=#2 \textwidth, angle=-90]{#1.eps}
        \hfill
        \includegraphics[width=#6 \textwidth, angle=-90]{#5.eps}
        \vskip 5 mm
        \parbox[t]{#4\textwidth}{\caption {#3}\label{#1}}
        \hfill
        \parbox[t]{#8\textwidth}{\caption {#7}\label{#5}}
    \end{center}
\end{figure} }

\def \myfigmis #1#2#3#4#5#6#7#8
{\begin{figure}[ht]
    \begin{center}
        \includegraphics[width=#2 \textwidth, angle=-90]{#1.eps}
        \hfill
        \includegraphics[width=#6 \textwidth]{#5.eps}
        \vskip 5 mm
        \parbox[t]{#4\textwidth}{\caption {#3}\label{#1}}
        \hfill
        \parbox[t]{#8\textwidth}{\caption {#7}\label{#5}}
    \end{center}
\end{figure} }

\def\myfigure #1#2#3#4
{\begin{figure}[ht]\begin{center}
\includegraphics[width=#2 \textwidth]{#1.eps}
\vskip 1 mm
\parbox[t]{#4\textwidth}{\caption{#3}\label{#1}}
\end{center}\end{figure}}

\def\myfig #1#2#3#4
{\begin{figure}[ht]\begin{center}
\includegraphics[width=#2 \textwidth, angle = -90]{#1.eps}
\vskip 1 mm
\parbox[t]{#4\textwidth}{\caption{#3}\label{#1}}
\end{center}\end{figure}}

\begin{document}
\title{Anisotropic cosmological models with a perfect fluid and a $\Lambda$ term}
\author{Bijan \surname{Saha}}
\affiliation{Laboratory of Information Technologies\\
Joint Institute for Nuclear Research, Dubna\\
141980 Dubna, Moscow region, Russia} \email{saha@thsun1.jinr.ru, bijan@jinr.ru}
\homepage{http://thsun1.jinr.ru/~saha/}
\date{\today}
\begin{abstract}
We consider a self-consistent system of Bianchi type-I (BI)
gravitational field and a binary mixture of perfect fluid and dark
energy given by a cosmological constant. The perfect fluid is
chosen to be the one obeying either the usual equation of state,
i.e., $p = \zeta \ve$, with $\zeta \in [0,\,1]$ or a van der Waals
equation of state. Role of the $\Lambda$ term in the evolution of
the BI Universe has been studied.
\end{abstract}



\maketitle

\bigskip

            \section{Introduction}

In view of its importance in explaining the observational
cosmology many authors have considered cosmological models with
dark energy. In a recent paper Kremer \cite{kremer} has modelled
the Universe as a binary mixture whose constitutes are described
by a van der Waals fluid and by a dark energy density. Zlatev {\it
et al.} \cite{zlatev} showed that "tracker field", a form of
qiuntessence, may explain the coincidence, adding new motivation
for the quintessence scenario. The fate of density perturbations
in a Universe dominated by the Chaplygin gas, which exhibit
negative pressure was studied by Fabris {\it et al.}
\cite{fabris}. Model with Chaplygin gas was also studied in the
Refs. \cite{dev,sen}. In doing so the author considered a
spatially flat, homogeneous and isotropic Universe described by a
Friedmann-Robertson-Walker (FRW) metric. Since the theoretical
arguments and recent experimental data support the existence of an
anisotropic phase that approaches an isotropic one, it makes sense
to consider the models of Universe with anisotropic back-ground in
presence of dark energy. The simplest of anisotropic models, which
nevertheless rather completely describe the anisotropic effects,
are Bianchi type-I (BI) homogeneous models whose spatial sections
are flat but the expansion or contraction rate is
direction-dependent. In a number of papers, e.g.,
\cite{PRD23501,PRD24010}, we have studied the role of a $\Lambda$
term in the evolution of a BI space-time in presence of spinor
and/or scalar field with a perfect fluid satisfying the equation
of state $p = \zeta \ve$. In this paper we study the evolution of
an initially anisotropic Universe given by a BI spacetime in
presence of a perfect fluid obeying not only $p = \zeta \ve$, but
also the Van der Waals equation of state.

            \section{Basic equations}
The Einstein field equation on account of the cosmological
constant we write in the form
\begin{equation}
R_\mu^\nu - \frac{1}{2} \delta_\mu^\nu R = \kappa T_\mu^\nu +
\delta_\mu^\nu \Lambda. \label{ein}
\end{equation}
Here $R_\mu^\nu$ is the Ricci tensor, $R$ is the Ricci scalar and
$\kappa$ is the Einstein gravitational constant. As was mentioned
earlier, $\Lambda$ is the cosmological constant. To allow a steady
state cosmological solution to the gravitational field equations
Einstein \cite{ein} introduced a fundamental constant, known as
cosmological constant or $\Lambda$ term, into the system. Soon
after E. Hubble had experimentally established that the Universe
is expanding, Einstein returned to the original form of his
equations citing his temporary modification of them as the biggest
blunder of his life. $\Lambda$ term made a temporary comeback in
the late 60's. Finally after the pioneer paper by A. Guth
\cite{guth} on inflationary cosmology researchers began to study
the models with $\Lambda$ term with growing interest. Note that in
our previous papers \cite{PRD23501,PRD24010} we studied the
Einstein field equations where the cosmological term appears with
a negative sign. Here following the original paper by Einstein and
one by Sahni \cite{sahni} we choose the sign to be positive. In
this paper a positive $\Lambda$ corresponds to the universal
repulsive force, while a negative one gives an additional
gravitational force. Note that a positive $\Lambda$ is often taken
to a form of dark energy.

We study the gravitational field given by an anisotropic Bianchi
type I (BI) cosmological model and choose it in the form:
\begin{equation}
ds^2 =  dt^2 - a^2 dx^2 - b^2 dy^2 - c^2 dz^2,
\label{BI}
\end{equation}
with the metric functions $a,\,b,\,c$ being the functions of time
$t$ only.

The Einstein field equations \eqref{ein} for the BI space-time in
presence of the $\Lambda$ term now we write in the form
\begin{subequations}
\label{ee}
\begin{eqnarray}
\frac{\ddot b}{b} +\frac{\ddot c}{c} + \frac{\dot b}{b}\frac{\dot
c}{c}&=&  \kappa T_{1}^{1} + \Lambda,\label{11}\\
\frac{\ddot c}{c} +\frac{\ddot a}{a} + \frac{\dot c}{c}\frac{\dot
a}{a}&=&  \kappa T_{2}^{2} +\Lambda,\label{22}\\
\frac{\ddot a}{a} +\frac{\ddot b}{b} + \frac{\dot a}{a}\frac{\dot
b}{b}&=&  \kappa T_{3}^{3} + \Lambda,\label{33}\\
\frac{\dot a}{a}\frac{\dot b}{b} +\frac{\dot b}{b}\frac{\dot c}{c}
+\frac{\dot c}{c}\frac{\dot a}{a}&=&  \kappa T_{0}^{0} + \Lambda.
\label{00}
\end{eqnarray}
\end{subequations}
Here over-dot means differentiation with respect to $t$. The
energy-momentum tensor of the source is given by
\begin{equation}
T_{\mu}^{\nu} = (\ve + p) u_\mu u^\nu - p \delta_\mu^\nu,
\label{emt}
\end{equation}
where $u^\mu$ is the flow vector satisfying
\begin{equation}
g_{\mu\nu} u^\mu u^\nu = 1.
\label{scprod}
\end{equation}
Here $\ve$ is the total energy density of a perfect fluid and/or
dark energy density, while $p$ is the corresponding pressure. $p$
and $\ve$ are related by an equation of state which will be
studied below in detail. In a co-moving system of coordinates from
\eqref{emt} one finds
\begin{equation}
T_0^0 = \ve, \qquad T_1^1 = T_2^2 = T_3^3 = - p.
\label{compemt}
\end{equation}
In view of \eqref{compemt} from \eqref{ee} one immediately obtains
\cite{PRD23501}
\begin{subequations}
\label{abc}
\begin{eqnarray}
a(t) &=&
D_{1} \tau^{1/3} \exp \bigl[X_1 \int\,\frac{dt}{\tau (t)} \bigr],
\label{a} \\
b(t) &=&
D_{2} \tau^{1/3} \exp \bigl[X_2 \int\,\frac{dt}{\tau (t)} \bigr],
\label{b}\\
c(t) &=&
D_{3} \tau^{1/3}\exp \bigl[X_3  \int\,\frac{dt}{\tau (t)} \bigr].
\label{c}
\end{eqnarray}
\end{subequations}
Here $D_i$ and $X_i$ are some arbitrary constants obeying
$$D_1 D_2 D_3 = 1, \qquad X_1 + X_2 + X_3 = 0,$$
and $\tau$ is a function of $t$ defined to be
\begin{equation}
\tau = a b c. \label{tau}
\end{equation}
From \eqref{ee} for $\tau$ one find
\begin{equation}
\frac{\ddot \tau}{\tau} = \frac{3 \kappa}{2} \bigl(\ve - p\bigr) +
3 \Lambda. \label{dtau}
\end{equation}
On the other hand the conservation law for the energy-momentum tensor gives
\begin{equation}
\dot{\ve} = -\frac{\dot \tau}{\tau} \bigl(\ve + p\bigr).
\label{dve}
\end{equation}
After a little manipulations from \eqref{dtau} and \eqref{dve} we
find
\begin{equation}
\dot{\tau}^2 = 3 (\kappa\ve + \Lambda) \tau^2 + C_1, \label{tve}
\end{equation}
with $c_1$ being an arbitrary constant. Let us now, in analogy
with Hubble constant, define
\begin{equation}
\frac{\dot \tau}{\tau} =  \frac{\dot a}{a} + \frac{\dot b}{b} +
\frac{\dot c}{c} = 3 H. \label{Hubble}
\end{equation}
On account of \eqref{Hubble} from \eqref{tve} one derives
\begin{equation}
\kappa \ve = 3 H^2 - \Lambda - C_1/(3 \tau^2). \label{tven}
\end{equation}
It should be noted that the energy density of the Universe is a
positive quantity. It is believed that at the early stage of
evolution when the volume scale $\tau$ was close to zero, the
energy density of the Universe was infinitely large. On the other
hand with the expansion of the Universe, i.e., with the increase
of $\tau$, the energy density $\ve$ decreases and an infinitely
large $\tau$ corresponds to a $\ve$ close to zero. Say at some
stage of evolution $\ve$ is too small to be ignored. In that case
from \eqref{tven} follows
\begin{equation}
3 H^2 - \Lambda \to 0.\label{limit}
\end{equation}
As it is seen  from \eqref{limit} in this case $\Lambda$ is
essentially non-negative. We can also conclude from \eqref{limit}
that in absence of a $\Lambda$ term beginning from some value of
$\tau$ the evolution of the Universe comes stand-still, i.e.,
$\tau$ becomes constant, since $H$ becomes trivial, whereas in
case of a positive $\Lambda$ the process of evolution of the
Universe never comes to a halt. Moreover it is believed that the
presence of the dark energy (which can be explained with a
positive $\Lambda$ as well) results in the accelerated expansion
of the Universe. As far as negative $\Lambda$ is concerned, its
presence imposes some restriction on $\ve$, namely, $\ve$ can
never be small enough to be ignored. It means in that case there
exists some upper limit for $\tau$ as well (note that $\tau$ is
essentially nonnegative, i.e. bound from below). In our previous
papers we came to the same conclusion \cite{PRD23501,PRD24010}
[with a positive $\Lambda$ which in the present paper appears to
be negative].

Inserting \eqref{Hubble} and \eqref{tven} into \eqref{dtau} one now finds
\begin{equation}
\dot H = - \frac{1}{2}\bigl(3 H^2 - \Lambda + \frac{C_1}{3 \tau^2}
+ \kappa p\bigr) = -\frac{\kappa}{2} \bigl(\ve + p\bigr) -
\frac{C_1}{3 \tau^2}\,. \label{Hdef}
\end{equation}
In view of \eqref{tven} from \eqref{Hdef} follows that if the
perfect fluid is given by a stiff matter where $p = \ve$, the
corresponding solution does not depend on the constant $C_1$.

Let us now go back to the Eq. \eqref{tve}. It is in fact the first
integral of \eqref{dtau} and can be written as
\begin{equation}
\dot \tau = \pm \sqrt{C_1 + 3(\kappa \ve + \Lambda) \tau^2}
\label{fi}
\end{equation}
On the other hand, rewriting \eqref{dve} in the form
\begin{equation}
\frac{\dot\ve}{\ve + p} = \frac{\dot \tau}{\tau},
\end{equation}
and taking into account that $p$ is a function of $\ve$, one
concludes that the right hand side of the Eq. \eqref{dtau} is a
function of $\tau$ only, i.e.,
\begin{equation}
\ddot \tau = \frac{3 \kappa}{2} \bigl(\ve - p\bigr) \tau + 3
\Lambda \tau = \mathcal F(\tau). \label{ddtau}
\end{equation}
From a mechanical point of view Eq. \eqref{ddtau} can be
interpreted as an equation of motion of a single particle with
unit mass under the force $\mathcal F(\tau)$. Then the following
first integral exists \cite{PRD24010}:
\begin{equation}
    \dot \tau = \sqrt{2[\mathcal E - \mathcal U(\tau)]}\,.
\label{1stint}
\end{equation}
Here $\mathcal E$ can be viewed as energy and $\mathcal U(\tau)$
is the potential of the force $\mathcal F$. Comparing the Eqs.
\eqref{fi} and \eqref{1stint} one finds $\mathcal E = C_1/2$ and
\begin{equation}
\mathcal U(\tau) = -\frac{3}{2}(\kappa \ve + \Lambda) \tau^2.
\label{poten}
\end{equation}
Let us finally write the solution to the Eq. \eqref{dtau} in
quadrature:
\begin{equation}
\frac{d\tau}{\sqrt{C_1 + 3(\kappa \ve + \Lambda) \tau^2}} = t +
t_0, \label{quad}
\end{equation}
where the integration constant $t_0$ can be taken to be zero,
since it only gives a shift in time.

In what follows we study the Eqs. \eqref{dtau} and \eqref{dve} for
perfect fluid obeying different equations of state.

\section{Universe filled with perfect fluid}
In this section we consider the case when the source field is
given by a perfect fluid. Here we study two possibilities: (i) the
energy density and the pressure of the perfect fluid are connected
by a linear equation of state; (ii) the equation of state is a
nonlinear (Van der Waals) one.

\subsection{Universe as a perfect fluid with $p_{\rm pf} =
\zeta \ve_{\rm pf}$}

In this subsection we consider the case when the source field is given
by a perfect fluid fluid obeying the equation of state
\begin{equation}
p_{\rm pf}\,=\,\zeta\,\ve_{\rm pf}, \label{eqst}
\end{equation}
where $\zeta$ is a constant and lies in the
interval $\zeta\, \in [0,\,1]$. Depending on its numerical value,
$\zeta$ describes the following types of Universes \cite{jacobs}
\begin{subequations}
\label{zeta}
\begin{eqnarray}
\zeta &=& 0, \qquad \qquad {\rm (dust\,\, Universe)},\\
\zeta &=& 1/3, \quad \qquad {\rm (radiation\,\, Universe)},\\
\zeta &\in& (1/3,\,1), \quad {\rm (hard\,\, Universes)},\\
\zeta &=& 1, \quad \qquad \quad {\rm (Zel'dovich\,\, Universe
\,\, or\,\, stiff\,\, matter)}.
\end{eqnarray}
\end{subequations}

In view of \eqref{eqst}, from \eqref{dve} for the energy density
and pressure one obtains
\begin{equation}
\ve_{\rm pf} = \ve_0/\tau^{(1+\zeta)}, \quad p_{\rm pf} = \zeta
\ve_0/\tau^{(1+\zeta)}, \label{vep}
\end{equation}
where $\ve_0$ is the constant of integration. For $\tau$ from
\eqref{quad} one finds
\begin{equation}
\frac{d\tau}{\sqrt{C_1 + 3(\kappa \ve_0 \tau^{1 - \zeta} + \Lambda
\tau^2)}} = t. \label{quadpf}
\end{equation}
As one sees, the positivity of the radical in \eqref{quadpf} for a
negative $\Lambda$ imposes some restriction on the upper value of
$\tau$, i.e., $\tau$ should be bound from above as well. In Fig.
\ref{potential} the graphical view of the potential $\mathcal
U(\tau)$ is illustrated for a negative $\Lambda$. As it was
mentioned earlier, $\mathcal E$ or $C_1$ in case of $\zeta = 1$
does not play any role. Universe in this case initially expands,
reaches to the maximum and then begin to contract finally giving
rise to a space-time singularity [cf Fig. \ref{oscillation}]. For
the other cases depending on the choice of $\mathcal E$ expansion
of the Universe is either non-periodic [Fig. \ref{nonperiodic}]
with a singularity at the end or oscillatory one without
space-time singularity [Fig. \ref{oscillation}]. In Fig.
\ref{inflation} we demonstrate the evolution of the BI Universe
with a positive $\Lambda$. In this case the Universe expands
exponentially, the initial anisotropy quickly dies away and the BI
Universe evolves into a isotropic FRW one. There does not any
upper bound for $\tau$ in case of a positive $\Lambda$. Note that
in the Figs. (\ref{potential} -  \ref{inflation})\,\, $\bf d$,\,\,
$\bf r$,\,\, $\bf h$ and $\bf s$ stand for dust, radiation, hard
Universe and stiff matter, respectively.

\vskip 1 cm

\myfigs{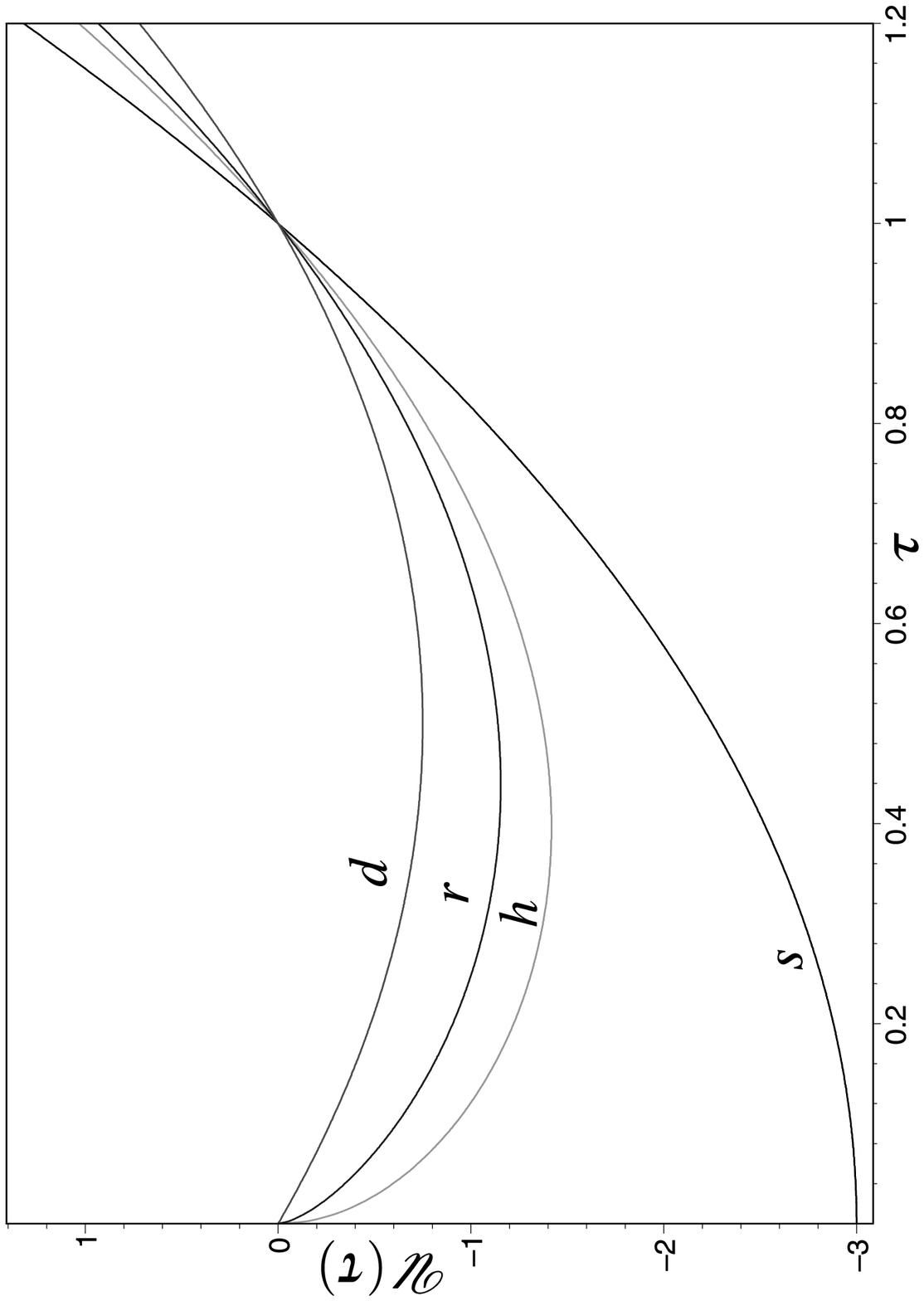}{0.32}{View of the potential $\mathcal U(\tau)$.
As one sees in case of stiff matter this potential allows only
non-periodic solution.}{0.45}{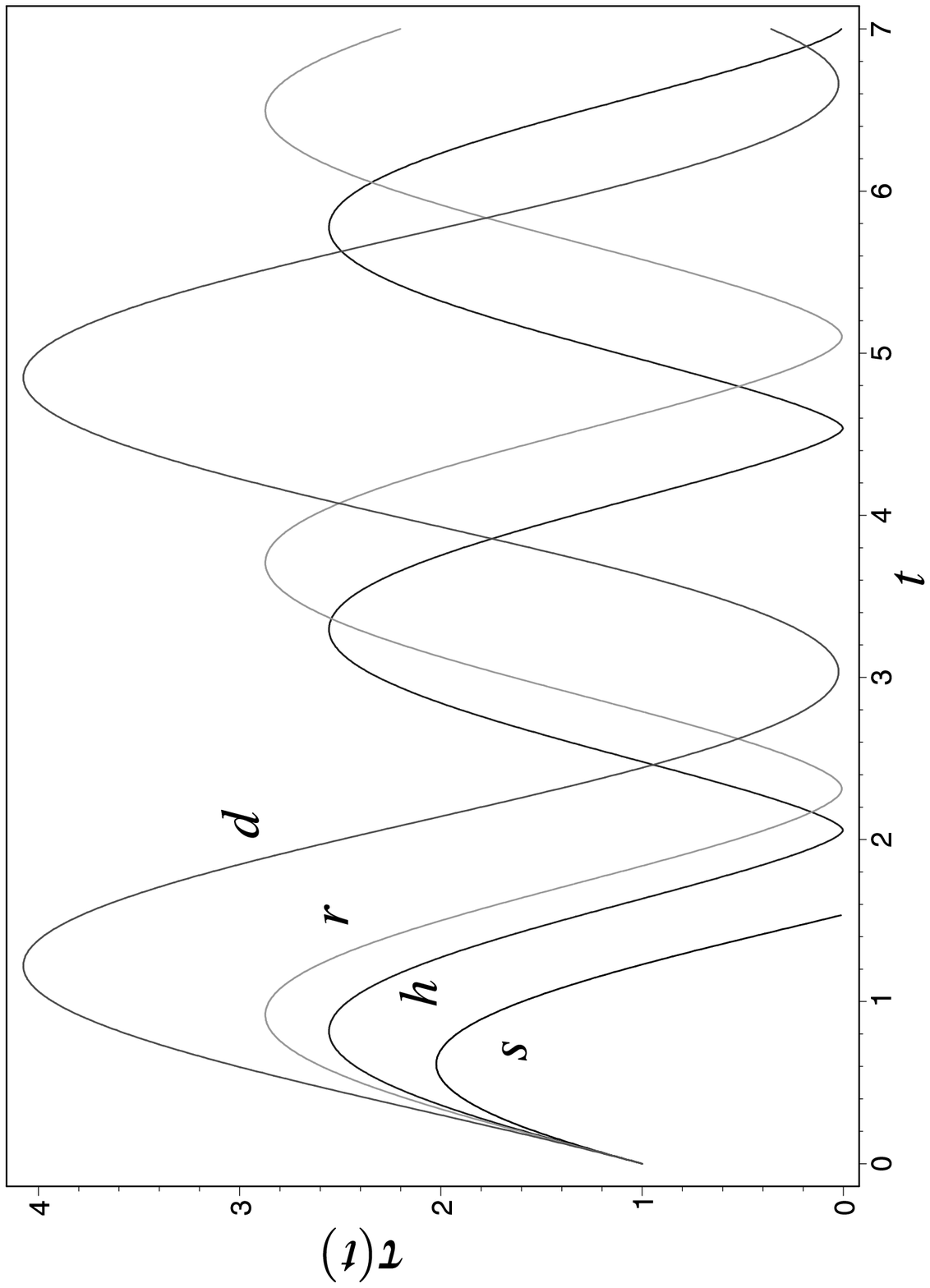}{0.30}{Evolution of
volume scale $\tau$ with a negative $\Lambda$ and $C = -0.1$. As
one sees, in this case the model with perfect fluid given by dust,
radiation and hard Universe allow oscillation, whereas, stiff
matter gives rise to a non-periodic solution.}{0.43}

\myfigs{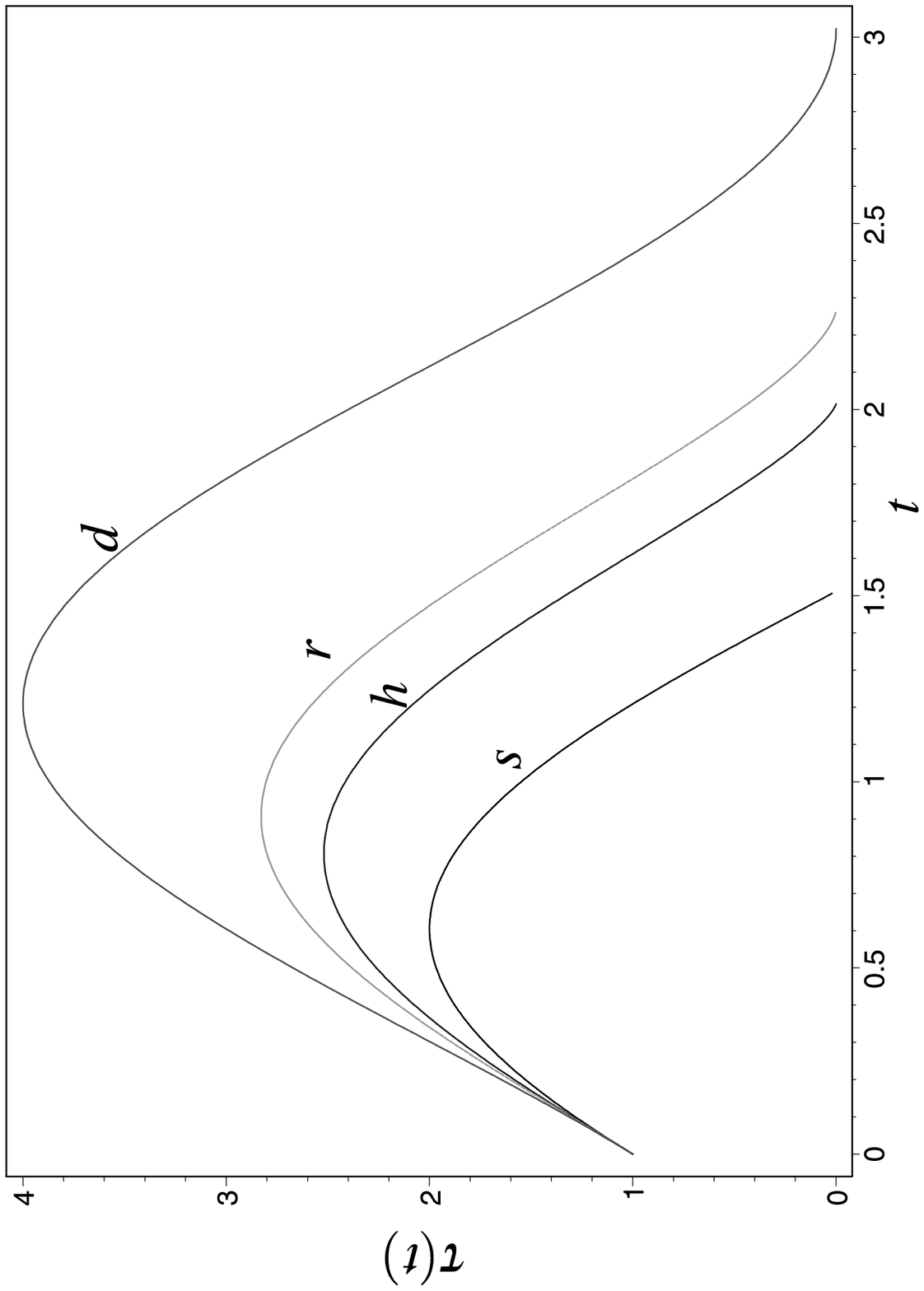}{0.30}{Evolution of volume scale $\tau$ with a
negative $\Lambda$ and $C = 0$. In this case the model with
perfect fluid given by dust, radiation, hard Universe and a stiff
matter gives rise to a non-periodic solution.}{0.45}
{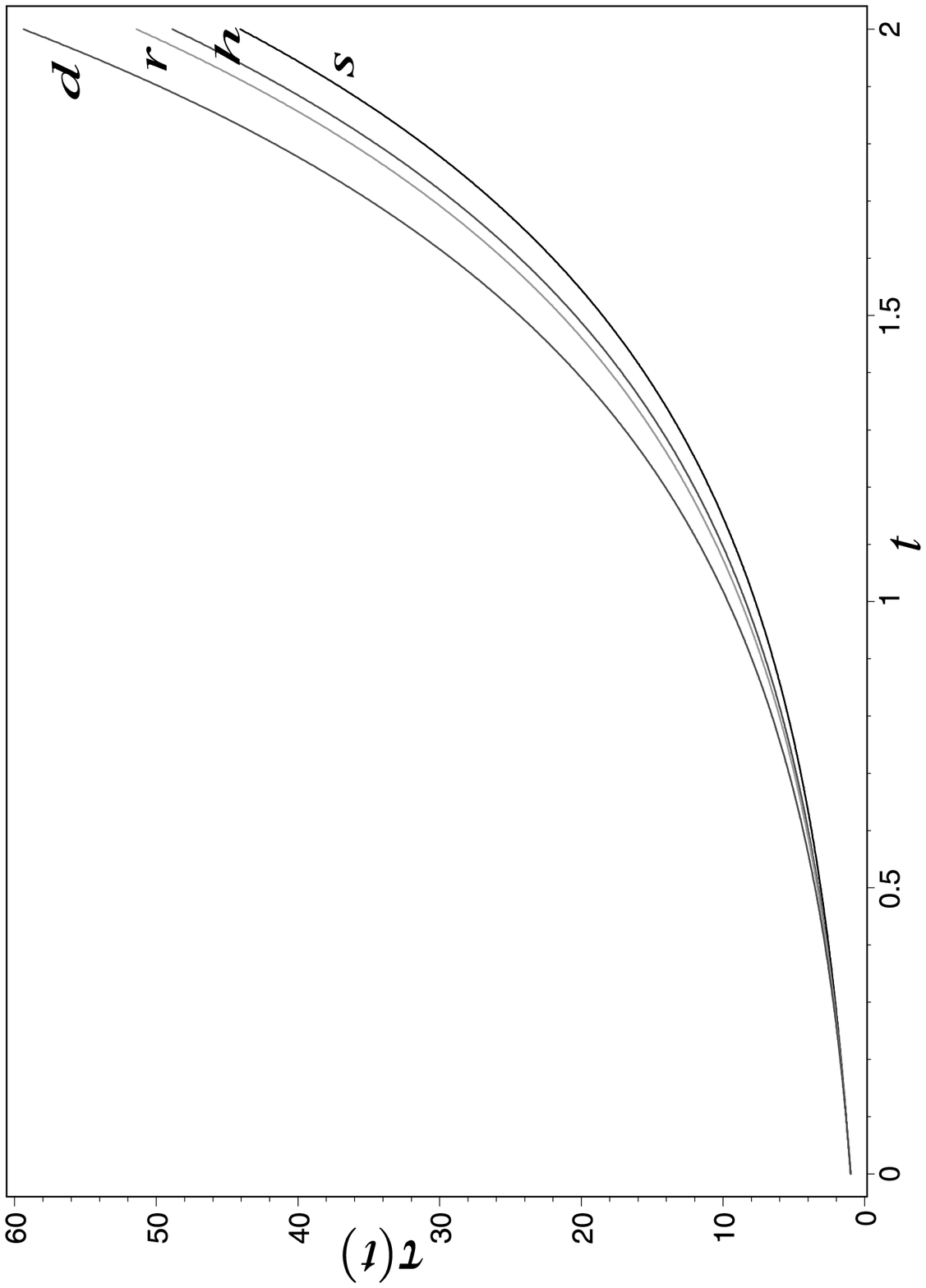}{0.30}{Evolution of the Universe with a positive
positive $\Lambda$. In this case independent to the choice of
$zeta$ the expansion of the Universe is always exponential.}{0.43}

In absence of the $\Lambda$ term one immediately finds
\begin{equation}
\tau = A t^{2/(1+\zeta)}, \label{tpf}
\end{equation}
with $A$ being some integration constant. As one sees from
\eqref{abc}, in absence of a $\Lambda$ term, for $\zeta < 1$ the
initially anisotropic Universe eventually evolves into an
isotropic FRW one, whereas, for $\zeta = 1$, i.e., in case of
stiff matter the isotropization does not take place.

\subsection{Universe as a van der Waals fluid}

Here we consider the case when the source field is given by a
perfect fluid with a van der Waals equation of state in absence of
dissipative process. The pressure of the van der Waals fluid
$p_{\rm w}$ is related to its energy density $\ve_{\rm w}$ by
\cite{kremer}
\begin{equation}
p_{\rm w} = \frac{8 W \ve_{\rm w}}{3 - \ve_{\rm w}} - 3 \ve_{\rm w}^2.
\label{wes}
\end{equation}
In \eqref{wes} the pressure and the energy density is written in
terms of dimensionless reduced variables and $W$ is a parameter
connected with a reduced temperature. In the Figs. \ref{enprLm1}
and \ref{enprLp1} the energy density and the pressure of the
system are illustrated with a negative and a positive $\Lambda$
term as well as in absence of it.

\myfigs{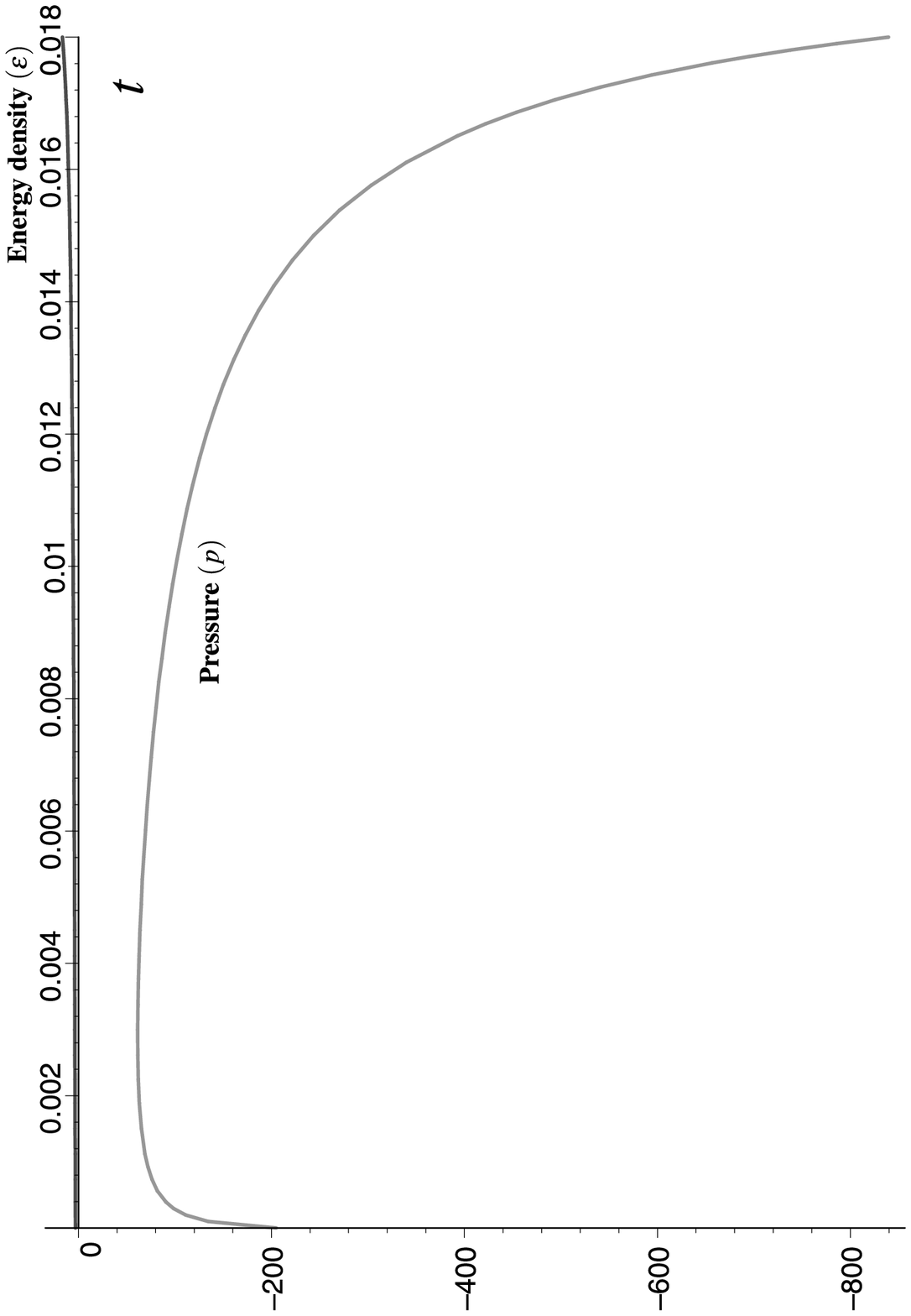}{0.30}{View of energy density $\ve$ and pressure
$p$ in case of a Van der Waals fluid with a negative
$\Lambda$.}{0.43}{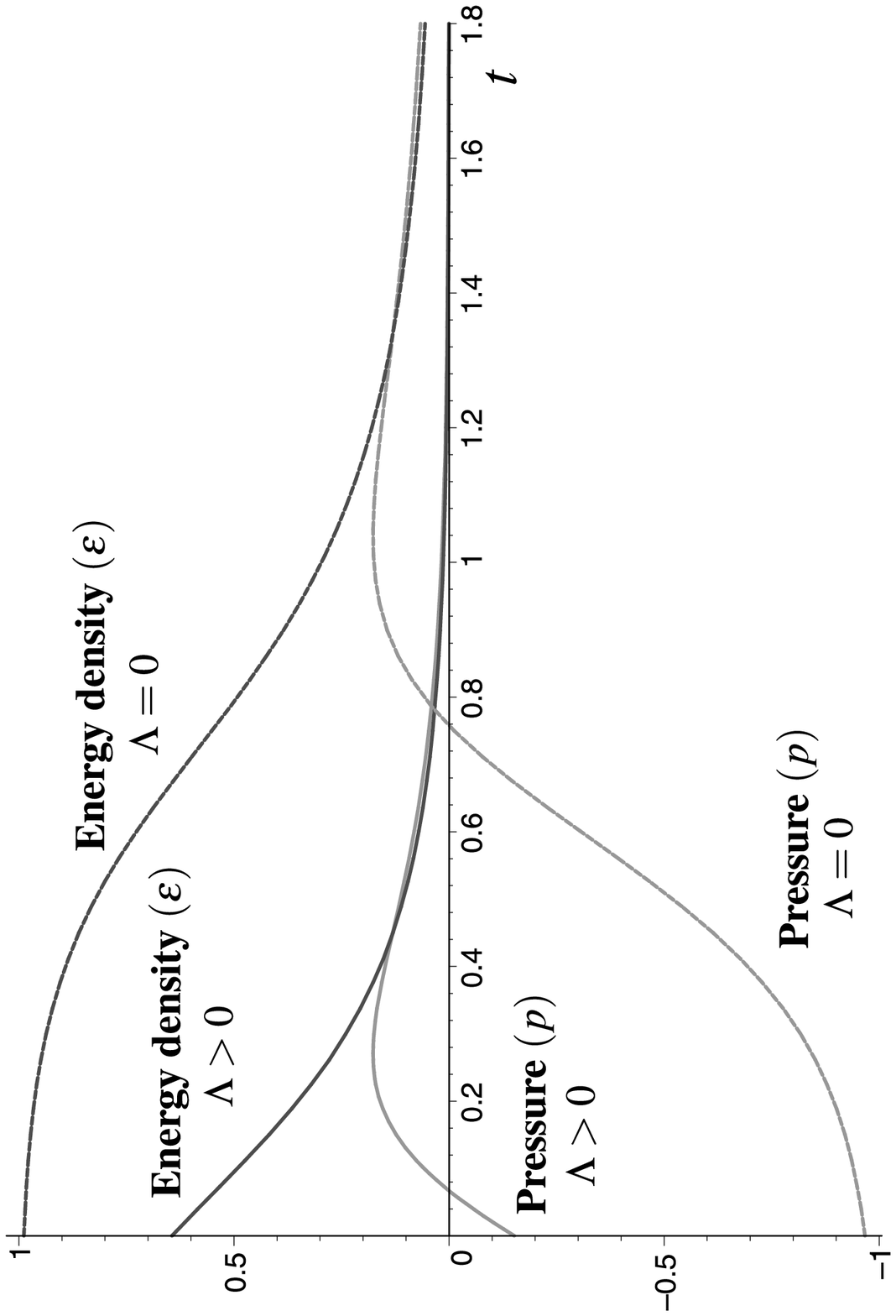}{0.30}{View of energy density $\ve$ and
pressure $p$ in case of Van der Waals fluid with $\Lambda \ge
0$.}{0.43}

\myfigs{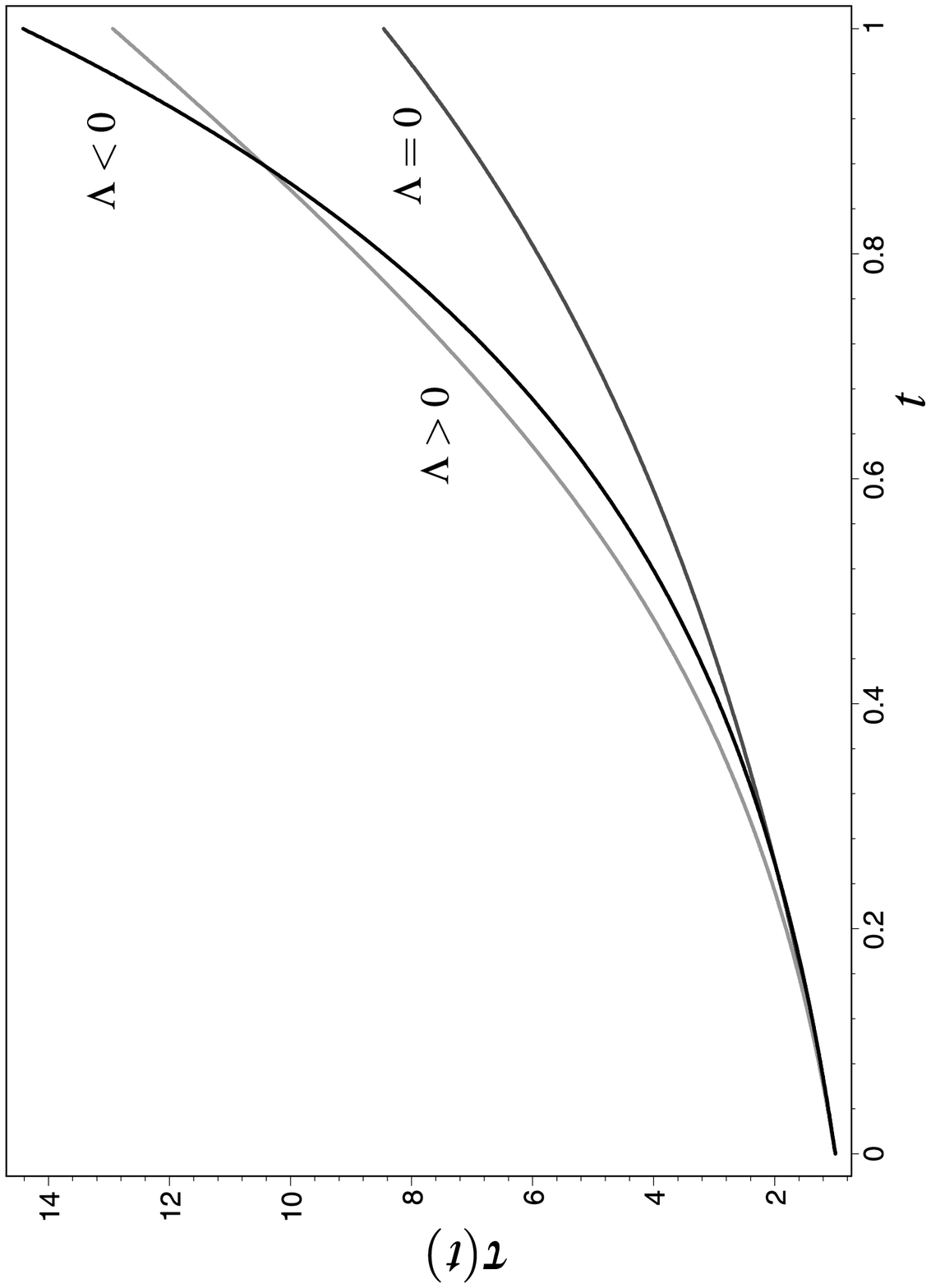}{0.28}{Evolution of $\tau (t)$ with the BI Universe
filled with Van der Waals fluid. Independent to the sign of
$\Lambda$ the model provides provides with exponentially expanding
Universe.}{0.45}{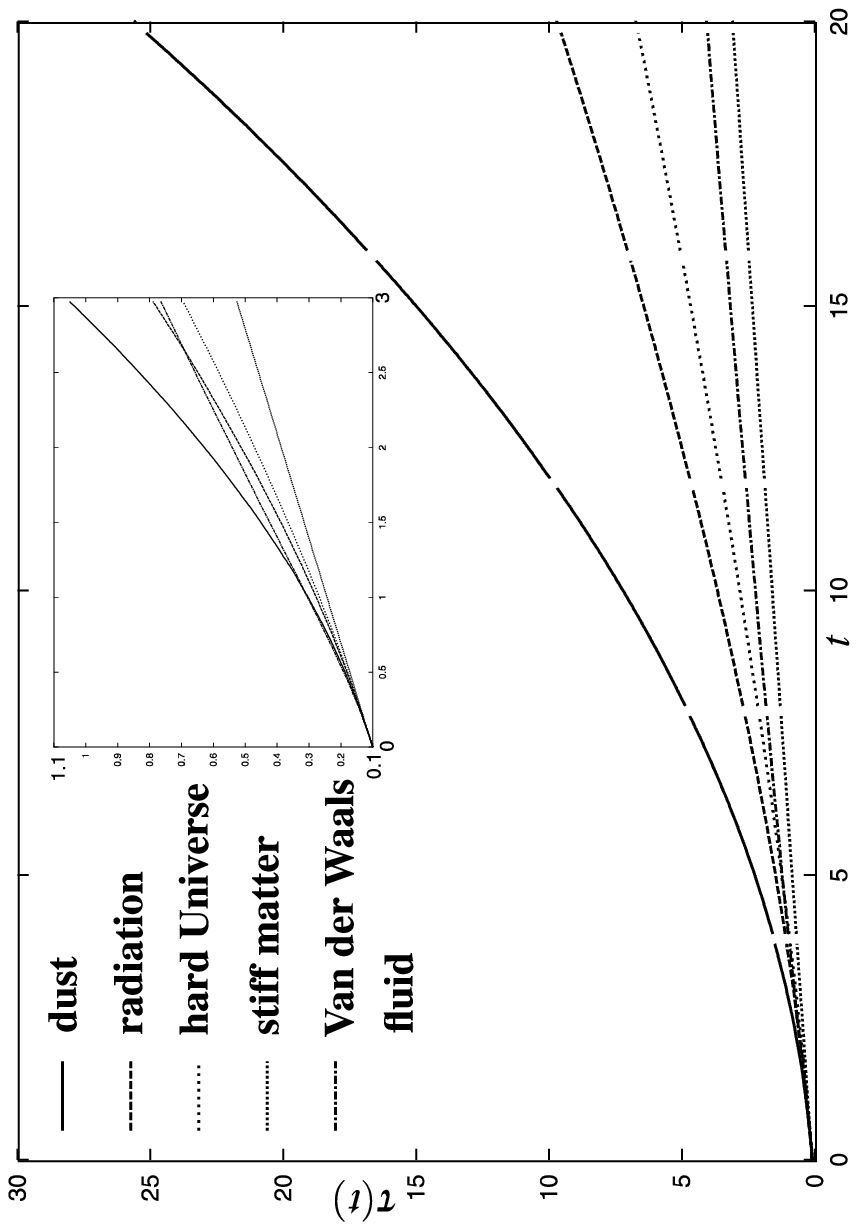}{0.31}{Evolution of the BI Universe filled
with different types of perfect fluid in absence of a $\Lambda$
term. As one sees, in case of Van der Waals fluid $\tau (t)$ grows
faster at the early stage, then slows down with time. }{0.45}

Inserting \eqref{wes} into \eqref{Hdef} on account of \eqref{tven}
one finds
\begin{eqnarray}
\dot{H} &=& -\frac{\{3 H^2 - \Lambda - C_1/(3 \tau^2)\}[(3 +
8W)\kappa -\{3 H^2 - \Lambda - C_1/(3 \tau^2)\}]}{2(3 \kappa - \{3
H^2 - \Lambda - C_1/(3 \tau^2)\})}\nonumber \\ &+&
\frac{3}{2\kappa}\bigl(\{3 H^2 - \Lambda - C_1/(3
\tau^2)\}\bigr)^2. \label{Hwes}
\end{eqnarray}
It can be easily verified that the Eq. \eqref{Hwes} in absence of
$\lambda$ term and $C_1 =0$ and $\kappa = 3$ coincides with that
given in \cite{kremer} :
\begin{equation}
\dot{H} = -\frac{3}{2}\Bigl[H^2 + \frac{8 W H^2}{3 - H^2} - 3
H^4\Bigr]. \label{Hwes1}
\end{equation}

The solution of the second-order differential equation
\eqref{Hwes} for $H(t)$ can be found by specifying the initial
value for $H(t)$ at $t = 0$, for a given value of parameter $W$.
Here we graphically present some results concerning the evolution
of BI Universe with a Van der Waals fluid. In Fig. \ref{infl1} we
compare the evolution of $\tau$ with and without $\Lambda$ term.
As one sees, the character of evolution does not depend on the
sign of $\Lambda$. In all cases we find exponentially expanding
Universe, though the rapidity of growth depends on $\Lambda$. The
Fig. \ref{Volume} gives the comparison of the expansion of $\tau$
with perfect fluid obeying different equations of state.

\section{Conclusion}

The evolution of an anisotropic Universe given by a Bianchi type I
cosmological model is studied in presence of a perfect fluid and a
$\Lambda$ term. It has been shown that in case of a perfect fluid
obeying $p = \zeta \ve$, where $p$ and $\ve$ are the pressure and
energy density of the fluid, respectively, a negative $\Lambda$
may generate an oscillation in the system thus giving rise to a
singularity-free mode of expansion. Introduction of a positive
$\Lambda$ in this case results in a rapid expansion of the
Universe. If the Universe is filled with a Van der Waals fluid, no
oscillatory or non-periodic mode of expansion occurs. Independent
to the sign of $\Lambda$ the Universe in this case expands
exponentially.

\newcommand{\hnl}{\htmladdnormallink}

\end{document}